\documentclass[12pt,preprint]{aastex}
\shorttitle{Globular Cluster Ages}
\shortauthors{Strader et al.}
\def\etal{{\it et al.}}
\input epsf

\begin{document}

\title{Extragalactic Globular Clusters: Old Spectroscopic Ages and New Views on Their Formation}

\author{Jay Strader, Jean P. Brodie, A. J. Cenarro, Michael A. Beasley}
\affil{UCO/Lick Observatory, University of California, Santa Cruz, CA 95064}
\email{strader@ucolick.org, brodie@ucolick.org, cenarro@ucolick.org, mbeasley@ucolick.org}

\author{Duncan A. Forbes}
\affil{Centre for Astrophysics \& Supercomputing, Swinburne University, Hawthorn VIC 3122, Australia}
\email{dforbes@astro.swin.edu.au}

\begin{abstract}

We present the results of a meta-analysis of Keck spectra of extragalactic globular clusters (GCs) in a sample of eight galaxies, ranging
from dwarfs to massive ellipticals. We infer ages for the metal-poor and metal-rich GCs in these galaxies through comparisons to Galactic
GCs. Both subpopulations appear to be no younger than their Galactic counterparts, with ages $\ga 10$ Gyr. This is the largest sample of
galaxies for which ages have been constrained spectroscopically. Our results support the formation of most GCs in massive galaxies at high
redshift. We propose a scenario for the formation of GC subpopulations that synthesizes aspects of both accretion and in situ approaches in
the context of galaxy formation through hierarchical merging.

\end{abstract}


\section{Introduction}

Bimodal color distributions were first discovered in extragalactic globular cluster (GC) systems in the early nineties (Ostrov, Geisler, \&
Forte 1993; Zepf \& Ashman 1993) and are now known to be commonplace among luminous galaxies (Larsen \etal~2001; Kundu \& Whitmore 2001). It
was realized early on that this finding had far-reaching implications for our understanding of the formation histories of GC systems. Since
GC formation is thought to accompany all major star forming episodes in a galaxy's history (Schweizer 2000), an understanding of the origin
of GC bimodality implicitly constrains the formation histories of their host galaxies.

Several scenarios have been developed to explain these GC subpopulations, which are generally termed metal-poor (MP) and metal-rich (MR). The 
major merger model (Schweizer 1987; Ashman \& Zepf 1992) predicts the formation of MR GCs out of enriched gas during spiral--spiral mergers 
that form elliptical galaxies. The MP GCs are associated with the progenitor spirals. Forbes, Brodie, \& Grillmair (1997) pointed out that 
bimodality could also arise in a multiphase in situ scenario: the MP GCs are formed in gaseous clumps within the early protogalactic potential, 
while the MR GCs are formed after a dormant period in a second phase of star formation. It is during this second phase that the bulk of the 
galaxy starlight is thought to be formed. Finally, C{\^ o}t{\' e} et al.~(1998, 2000, 2002) suggested an accretion scenario in which a massive 
seed galaxy forms MR GCs in situ, while MP GCs are acquired via the accretion of dwarf galaxies and their accompanying low-metallicity GC 
systems. GC formation has also recently been simulated with semi-analytic models (SAMs; Beasley \etal~2002) and N-body+gasdynamic codes 
(Kravtsov \& Gnedin 2005). The Beasley \etal~model, which followed GC systems to the present day, found it necessary to truncate MP GC 
formation at $z > 5$ to produce bimodality. Candidates for this truncation process include feedback in low-mass halos (Forbes \etal~1997) and 
cosmic reionization (Santos 2003).

Some of the scenarios mentioned above were developed before the clear success of $\Lambda$CDM cosmology and the hierarchical merging
paradigm. However, they are relevant for understanding structure formation. These scenarios describe the behavior of baryons within dark matter
halos, which is dominated by processes that are relatively unconstrained at present. All of the scenarios predict old ages for MP GCs, but
not necessarily for MR GCs. In the merger model, the mergers (and hence the MR GCs) will have a range of ages. In the multiphase scenario
both GC subpopulations are expected to be old ($\ga 10$ Gyr) with the MR GCs slightly younger than the MP GCs. Some variation with
environment and galaxy mass may be expected. In the accretion scenario, both subpopulations are $\sim$ coeval and very old. Of course,
reality is likely to be some hybrid of the these alternatives, and in a very real sense the distinctions between them blur if the processes
occur at high enough redshift. Nonetheless, it is important to determine whether one (or more) of these ideas represents a dominant
mechanism. Thus, the goal of age-dating GC subpopulations and, in particular, detecting or ruling out age differences between MR and MP GCs,
will set constraints on the processes of galaxy formation and limits on the epochs of galaxy assembly.

The goal of this work is to determine relative ages of GC subpopulations and set limits on their absolute ages (modulo modeling and
observational uncertainties). Our data are among the best spectra currently available for extragalactic GCs, and it is unlikely that
significantly tighter age constraints for GCs will be forthcoming using Lick index diagnostics (Trager \etal~1998 and references therein) in
the era of $8-10$ m class telescopes. Since the Lick system is not optimized for studying stellar systems with low velocity dispersion, this
current short study is intended as a summary of our spectroscopic work in age-dating GCs and as a starting point for our future
investigations in which new, higher resolution (3 \AA) index definitions will be employed.

\section{The Sample and Method}

Not all galaxies for which our group has previously obtained spectra are included in this study. Here we require all Lick indices (see details
below) from H$\delta_F$ to Fe5335 to be measurable, and that at least two GCs per subpopulation have $H\beta$ and H$\delta_F$ errors $< 0.5$ \AA.
This left a sample of eight galaxies, which are listed in Table 1. The sample is dominated by massive early-type galaxies in group and cluster
environments, but includes two dwarfs.

The GC spectra were obtained at Keck with LRIS (Oke \etal~1995) over a variety of runs, from Dec.~1995 to Apr.~2004. Spectra were taken in
multislit mode (except for the GCs of the Fornax dwarf spheroidal), with resolution ranging from 3.3--8 \AA. The spectral coverage was
typically $\sim 4000-5500$ \AA, but varied with slit placement and from run to run, sometimes going further to the red or blue.

We refer the reader to the respective papers noted in Table 1 for details on the reduction of each data set; however, a short summary
follows. GC spectra extracted from individual exposures (typically 30 min in length) were wavelength calibrated using arc lamps and then
co-added. Small zero-point shifts to the wavelength calibration were made using bright night sky lines, and a standard flux calibration was
carried out using one or two standards from Massey \etal~(1988).

Lick indices were measured using the definitions in Trager \etal~(1998) and Worthey \& Ottaviani (1997). The spectra were smoothed to the
wavelength-dependent resolution (8--11 \AA) of the Lick system before the measurement of indices. An unfortunate aspect of the present study
is that we cannot fully calibrate our index measurements onto the Lick system. The number and type of Lick standard stars varied from run to
run, rendering consistent corrections impossible. We note that many of our previous studies of GCs in individual galaxies (e.g., Brodie
\etal~2005; Strader, Brodie, \& Forbes 2004a) have found such corrections unnecessary: they are generally smaller than the error bars on
individual index measurements, and there appear to be no large systematic effects. However, our lack of Lick corrections could still result
in small offsets when comparing to Lick index predictions of simple stellar population (SSP) models.

For this reason, we have decided to concentrate on a \emph{differential} analysis between our data and integrated spectra of Galactic GCs. 
These Galactic data are from Gregg (1994) and Puzia \etal~(2002b). In the former case we measured indices directly from the digital spectra, as 
described above. In the latter case we ``de-corrected" the published indices using Lick offsets given in the same paper. The resulting datasets 
should be amenable to consistent comparisons.

Even with our best individual spectra, estimating accurate ages and metallicities is difficult. An error in H$\beta$ of only $\sim 0.3$ \AA\ 
corresponds to an age error of $\sim 3$--4 Gyr for a 13 Gyr old GC. Thus, to facilitate comparison to the Galactic GCs, we constructed 
\emph{mean} MP and MR (and in NGC 4365 and NGC 1407, intermediate-metallicity) subpopulations for all galaxies, except for the two dwarfs 
Fornax and VCC 1087. Fornax has only MP GCs; VCC 1087 may have a small number of MR GCs, but they were sparsely sampled in our spectroscopy and 
do not meet the criteria outlined above (Beasley \etal~2005). Photometric studies of the remainder of our sample galaxies show clear evidence 
for subpopulations, so this division is justified (see individual papers for details). GCs were sorted into subpopulations primarily on the 
basis of this photometry (with typical separation at $V-I = 1.05$, $B-I = 1.85$, or $g-z = 1.1$). These criteria were augmented by principle 
components analysis (PCA)-based metallicities when ambiguous (Strader \& Brodie 2004). The number of GCs per subpopulation are listed in Table 
1. Lick indices for these subpopulations are weighted means of the individual GC indices with $2 \sigma$ rejection. The variance is calculated 
as the sum of the weights and is equivalent to a weighted standard variance of the mean. The mean subpopulation indices and standard errors of 
the mean are given in Tables 2 and 3. The only GCs excluded outright were two suspected young ($\sim 1$--2 Gyr) metal-rich GCs in the 
dynamically young elliptical NGC 3610 (Strader \etal~2003a; Strader \etal~2004a), since our focus is on deriving the properties of the ``old"  
subpopulations. For each subpopulation with at least three GCs, weighted standard deviations of the indices are given in Table 4, and represent 
the combined effects of observational error and intrinsic differences among the GCs within each subpopulation.

The majority of our GC samples are biased. First there is the necessary luminosity bias: nearly all of our GCs have $V < 23$. Adequate S/N 
cannot be achieved for GC fainter than this, even in $\sim 8-10$ hours at Keck. This apparent magnitude corresponds to $M_V \sim -8$ at the 
distance of Virgo, 0.5 mag brighter than the turnover of the GC luminosity function (LF). However, neither metallicity or age correlates 
strongly with GC mass in the Milky Way, suggesting that this introduces minimal bias in derived properties. More importantly, the GCs were not 
selected uniformly in color. This makes it impossible to construct ``true'' composite subpopulations that accurately reflect the underlying 
metallicities. However, we \emph{can} constrain mean ages, since younger GC subpopulations with a standard log-normal LF will be, on average, 
brighter than their old counterparts, even taking into account their probable higher metallicities. Thus magnitude-selected samples will give 
lower limits to mean GC subpopulation ages, unless the LF of the younger GCs is biased to low masses, either at birth or because of differing 
GC destruction.

\section{Ages}

The traditional method of estimating ages and metallicities using Lick indices involves constructing index-index plots. [MgFe], the harmonic 
mean of Mg$b$ and (Fe5270+Fe5335)/2, is often used as a metallicity indicator largely insensitive to variations in [$\alpha$/Fe]. However, the 
use of [MgFe] does not take advantage of the metallicity information contained in the remainder of the Lick indices. We have developed a 
PCA-based method for estimating GC metallicities using 11 Lick indices (Strader \& Brodie 2004), and choose to use these more robust 
metallicities instead of the standard [MgFe] (or the slightly different [MgFe]$\arcmin$, in which $<$Fe$>$ is replaced by ($0.72 \times$ Fe5270 
+ $0.28 \times$ Fe5335); Thomas, Maraston, \& Bender 2003).

Puzia \etal~(2004) have argued that the best Lick Balmer index for estimating GC ages is H$\gamma_F$. However, the blue pseudocontinuum 
bandpass of H$\gamma_F$ encompasses most of the CH absorption of the G band centered at $\sim 4300$ \AA. This makes the index extremely 
sensitive to changes in the carbon abundance (and overall metallicity, though its sensitivity decreases toward high metallicities). Thus, we 
have chosen not to use either of the Lick H$\gamma$ indices in this study. As noted above, we are currently engaged in a redefinition of the 
Lick system that is optimized for GCs.

Instead, we will use the average of H$\beta$ and $H\delta_F$ as our age-sensitive Balmer indices (in fact, using H$\gamma_F$ in the average as 
well gives consistent results). In Figure 1 we plot PCA [m/H] vs.~(H$\beta $+$H\delta_F$)/2. The formal random errors for PCA [m/H] are 
negligible, but we assign a systematic calibration error of 0.12 dex to all points (see Strader \& Brodie 2004). Our GC subpopulations fall at 
or below the envelope defined by the Galactic GCs, suggesting that they are no younger than Galactic GCs at all metallicities. 14 Gyr model 
lines for [$Z$/H] = $-2.25$ to 0 and [$\alpha$/Fe] = 0 and +0.3 from Thomas, Maraston, \& Korn (2004, TMK) are superposed. These include a blue 
horizontal branch below [$Z$/H] = $-1.35$. The models are plotted only to show that our index combinations are relatively insensitive to 
[$\alpha$/Fe] variations; we cannot make any direct model comparisons since our data are not on the Lick system. Indeed, ages derived using PCA 
metallicities are even less sensitive to [$\alpha$/Fe] than those derived from [MgFe]$\arcmin$. Thus, the use of PCA metallicities is 
preferable when deriving SSP ages for GCs, and this is likely to be true even for more metal-rich stellar systems, up to the current PCA 
calibration limit of solar metallicity. The MR Galactic GC with high Balmer line strength is NGC 6553; its location is due not to youth or warm 
horizontal branch stars but probably poor sampling in the crowded bulge (Puzia \etal~2002b).

Let us consider the scatter among the mean subpopulations in Figure 1 more quantitatively. The relation between [m/H] and Balmer line strength 
appears to be linear to solar metallicity; a least-squares fit to all points with [m/H] $< 0$ gives: (H$\beta$+$H\delta_F$)/2 = $-1.221$ \, 
[m/H] + 0.632, with a residual standard error of 0.17. The mean Balmer error for these subpopulations is 0.16. Thus, the extragalactic points 
are entirely consistent with expected scatter in the mean values around a single line. This need not necessarily be a single isochrone (as 
there could be a shift to younger ages with increasing metallicity), but it does indicate that there is not a large age spread from galaxy to 
galaxy at fixed metallicity.

What can we conclude about the absolute ages of our GC subpopulations? Our results are directly tied to knowledge of the ages of MP and MR GCs 
in the Milky Way (MW). Unfortunately, there is considerable uncertainty in this subject as well. Using $\alpha$-enhanced isochrones with 
diffusion, Gratton \etal~(2003) find that the oldest MP GCs have formal ages of 13.4 $\pm 0.8 \pm 0.7$ Gyr (random and systematic errors). 
Krauss \& Chaboyer (2002) used a Monte Carlo technique to perform isochrone fits to simulated metal-poor color-magnitude diagrams with varying 
physical parameters, and found best-fit ages of 12.6 Gyr with $1 \sigma$ errors of $\sim 1$ Gyr. A recent revision of the 
$^{14}$N($p,\gamma$)$^{15}$O reaction rate raises these ages by 0.7--1.0 Gyr (Imbriani \etal~2004); they are indistinguishable from the age of 
the universe within the errors. Similar ages are found through integrated spectroscopy (Proctor, Forbes, \& Beasley 2004).

Few MR GCs have robust (determined through direct main-sequence fitting) age determinations. Even for the best-studied of these GCs, 47 Tuc, 
there are still significant discrepancies, depending on the choice of subdwarfs and colors used for fitting. 47 Tuc could be coeval with MP 
GCs, or $\sim 2-3$ Gyr younger at 11 Gyr (Gratton \etal~2003, Grundahl, Stetson, \& Andersen 2002; or $\sim 12$ Gyr with the updated 
$^{14}$N($p,\gamma$)$^{15}$O rate). The main source of error is the distance modulus, and a convincing resolution of the issue may require 
direct parallax measurements from SIM or GAIA. In addition, though the bulk of the MR GCs are probably associated with the bulge (Minniti 1995; 
C{\^ o}t{\' e} 1999; Forbes, Larsen, \& Brodie 2001), a subset may belong to the thick disk---including 47 Tuc. Such thick disk clusters, if 
they exist, probably have few counterparts in early-type galaxies. Secure turnoff ages for more MR GCs will be needed before a more certain 
estimate of their mean properties can be made, but for the purposes of this paper we assume a minimum age of 10 Gyr for MR GCs.

In conclusion, the GC subpopulations in our sample all appear to have mean ages $\ga 10$ Gyr. However, relative \emph{mean} subpopulation ages 
more accurate than $\sim 1-2$ Gyr ($1 \sigma$) cannot be determined at present, and age errors on individual clusters are $\ga 3-4$ Gyr.

\section{Discussion}

The old GC ages derived in this paper are consistent with an emerging observational and theoretical consensus that very massive galaxies,
especially in dense environments, formed the bulk of their stars at high redshift (McCarthy \etal~2004; Chen \& Marzke 2004; Somerville 2004;
Nagamine \etal~2001). There is no sharp cutoff for the classification of a galaxy as ``very massive'', but a number often adopted is 3$\,
L^\star$ = $2 \times 10^{11} M_\odot$, which corresponds to $M_V \sim -21.5$ for an old metal-rich stellar population. Most spectroscopic
studies of GC systems beyond the Local Group have focused on such galaxies, since they generally have the largest GC systems and thus are
easiest to study. Our findings in this paper, coupled with the enormous number of photometric studies of GCs in such galaxies over the last
several decades, are entirely consistent with the formation at high redshift of most GCs in massive galaxies in dense environments. These ages
are consistent with previous spectroscopic studies of GC systems in individual galaxies by our group and by other workers (e.g., Kuntschner
\etal~2002; Cohen, Blakeslee, \& Ryzhov 1998).

Old ages for the majority of GCs in galaxies of all types might seem, on the face of it, to conflict with the discovery of young
``proto-GC'' candidates in gas-rich mergers (e.g., NGC 1275, Holtzman \etal~1992; NGC 7252, Whitmore \etal~1993; the Antennae, Whitmore \&
Schweizer 1995; NGC 3921, Schweizer \etal~1996) and intermediate-age ($\sim 1$--5 Gyr) GCs in some merger remnants. These latter systems have
been targeted because they have already suffered the bulk of their dynamical evolution and can thus be more readily compared to old GC systems.
There are only two galaxies for which intermediate-age GCs have been spectroscopically confirmed: NGC 1316 (Goudfrooij \etal~2001) and NGC 3610
(Strader \etal~2003a), though the existence of such GCs has been suggested on the basis of lower S/N spectra in NGC 5128 (Peng, Ford, \&
Freeman 2004) and NGC 1399 (Forbes \etal~2001). NGC 1316 appears to have a significant population of intermediate-age clusters, as revealed by
variations of the GCLF with galactocentric radius (Goudfrooij \etal~2004). However, these clusters make up only a minor component of the total
GC system in NGC 3610 (Strader \etal~2004a). The 1--2 Gyr merger remnant NGC 1052 shows no evidence for intermediate-age GCs (Pierce
\etal~2005). On the basis of optical/near-IR photometry, it has been suggested that intermediate-age subpopulations exist in several other
galaxies (e.g., NGC 4365, Puzia \etal~2002a; NGC 5846, Hempel \etal~2003). Recent high S/N, follow-up spectroscopy of prime intermediate-age
candidates in NGC 4365 found only old GCs (Brodie \etal~2005), showing the need for high S/N spectroscopic confirmation. The 
intermediate-age merger remnants studied thus far are a mixed bag; at least some remnants have smaller populations of intermediate-age GCs than 
might be expected if the major merger model is correct.

In addition to GC ages, a successful formation scenario must also reproduce the GC metallicity--galaxy mass relations for \emph{both}
MP and MR GCs. The slopes of these relations are now well-determined (Larsen \etal~2001, Strader, Brodie, \& Forbes 2004b),
and can offer important quantitative constraints on GC formation models, although they have not yet been used in this respect. The relations
are also useful for defining qualitative scenarios for the formation of massive ellipticals, as discussed below.

The in situ scenario (Forbes, Brodie \& Grillmair~1997) is sometimes considered together with the SAM of Beasley \etal~(2002) under the rubric
``hierarchical". In some sense all modern scenarios must be hierarchical, given the strength of evidence for the assembly of present-day DM
halos through hierarchical merging in a $\Lambda$CDM cosmology. Indeed, the accretion models of C{\^ o}t{\' e} et al. (1998, 2000, 2002)
were explicitly inspired by the convergence toward hierarchical galaxy assembly. While this specific scenario is difficult to reconcile with
recent observations (since the mean metallicities of GCs in dwarf galaxies are lower than those of MP GCs in massive galaxies; Strader
\etal~2004b), the hierarchical nature of structure formation implies that all models are accretion models to a certain extent.

\subsection{Metal-Poor Globular Clusters}

A key problem with the accretion model (and a point not considered in Strader \etal~2004b) is that local dwarfs cannot be directly compared
to their counterparts at high redshift (see, e.g., Santos 2003 as it relates to MP GCs, and Robertson \etal~2005 on the formation of the
Galactic halo). Dwarf galaxies sitting ``on top'' of a large overdensity destined to become a massive galaxy will collapse before those on
the periphery. If MP GC formation was terminated by reionization, the dwarfs that collapsed first might have produced MP GCs of a higher
metallicity than those dwarfs that collapsed later.

This could happen in a variety of ways: the dwarfs collapsing first could accrete a larger quantity of pristine gas (thereby enhancing star
formation), have more time to process the gas they accreted, or capture more enriched, outflowing gas from the higher density of nearby starforming
halos. These more centrally concentrated halos would generally be accreted into their parent halo, with their GCs forming its MP subpopulation,
while some of the outlying halos would survive to be present-day dwarfs. This scenario could qualitatively reproduce the MP GC-metallicity--galaxy
mass relation, since larger overdensities collapse first and thus would have MP GCs of higher metallicity. Since the MP GCs are accreted from
low-mass halos that are enriched ``in place" within a DM overdensity, our picture can be seen as a synthesis of the accretion and in situ
approaches. We have begun work to test this hypothesis using SAM+N-body halo merger trees and simple chemical evolution models. Recent numerical
simulations of the assembly of Milky Way-sized halos (Moore \etal~2005, priv.~communication) indicate that the radial distribution of MP Galactic
GCs they predict is consistent with our scenario.

We also note here the continuing efforts to place GC formation in a cosmological context. Given the lack of observational evidence for DM
halos around GCs (Moore 1996; Odenkirchen \etal~2003), a popular view among cosmologists is that MP GCs formed with halos that were later
stripped away (e.g., Cen 2001; Bromm \& Clarke 2002; Scannapieco et al.~2004) or are still present but difficult to detect (Mashchenko \&
Sills 2005). However, any model which requires different formation mechanisms for MP and MR GCs must contend with the strong similarities
between the two subpopulations, including sizes and the shapes of the mass and metallicity distributions (see also Harris 2003).

What can we infer about the masses of halos in which MP GCs formed? A number of Galactic satellites with total masses $\sim 10^6$--$10^7
M_{\odot}$ lack GCs entirely (Forbes \etal~2000), while the Fornax and Sagittarius dSphs, with masses $\ga 10^8 M_{\odot}$ (Walcher
\etal~2003; Law \etal~2005), have at least five GCs each. The dSph DDO 78 in the M81 group is slightly less massive and has one GC (Sharina
\etal~2003). Grebel \etal~(2000) find a candidate GC of the M31 satellite And I ($M_V \sim -12$; $M_{tot} \sim$ a few $\times 10^7$ for an
$M/L$ similar to Fornax), but at a projected distance of $\sim 40$ kpc from M31 the GC could be a contaminant. These results might suggest
that MP GCs formed in halos with a minimum mass of $\sim 10^7$--$10^8 M_{\odot}$. This ``intermediate'' regime could be better probed with
further study of And I (and the similar mass satellite And II). 

\subsection{Metal-rich Globular Clusters and Formation Efficiency}

Despite a variety of evidence suggesting that massive, compact star clusters form in most (possibly all) major episodes of star
formation (e.g., Schweizer 2000), and that a subset of these clusters may survive to become GCs, the relation between GCs and field
stars is poorly understood. For lack of data, arguments for the universality of GC formation efficiency (e.g., McLaughlin 1999) have
generally not been based on the separation of GC and field star subpopulations. In the MW, modern interpretations associate MP and
MR GCs with the halo ($\sim 1.5 \times 10^9 M_{\odot}$; Carney, Latham, \& Laird 1990) and bulge ($\sim 10^{10} M_{\odot}$),
respectively. There are $\sim 100$ MP and 50 MR GCs in the MW. The subpopulations have mass-normalized specific frequencies ($T$ =
the number of GCs per $10^9$ $M_{\odot}$; Zepf \& Ashman 1993) of 67 and 5, respectively. The ``universal'' GC formation efficiency
of McLaughlin corresponds roughly to $T \sim 8$ (assuming all gas in a system has been converted to stars), and is rather similar
to the MR value. However, the MP value is more than a factor of 10 higher than the MR one. 

A strong link between MR GCs and field stars is predicted in all the GC formation scenarios discussed to date and is consistent with observations
of several extragalactic GC systems (e.g., M31: Jablonka \etal~2000; NGC 5128: Harris \etal~1999, Harris \& Harris 2002; NGC 1399: Forte, Faifer,
\& Geisler 2005) and the nearly-constant ``bulge specific frequency" in elliptical galaxies and many spiral bulges (Forbes, Brodie \& Larsen~2001;
see Goudfrooij \etal~2003 for a slightly different view). In none of these galaxies have strong constraints been placed on the existence of a
population of metal-poor ``halo'' field stars; existing surveys in NGC 5128 and M31 (though see Guhathakurta \etal~2005) have found few metal-poor
stars, suggesting that the MP $T$ value may generically be higher than the MR. In our scenario, truncation by reionization offers a natural
explanation for these high $T$ values of MP GC systems.

The intimate connection between MR GCs and the underlying galaxy starlight has implications for the present study, as the old ages
of MR GCs in our sample of galaxies suggests old ages for the bulk of their field stars. However, with our small samples of GCs, we
cannot constrain at present whether they were formed in a short-lived ``monolithic'' dissipational event nor whether a small age
spread ($< 2$ Gyr) might exist.

Both in situ and accretion scenarios predict old ages for red GCs (but not the formation mechanism). The classical major merger picture (Ashman 
\& Zepf 1992), however, would predict younger ages, unless the main epoch of major mergers is pushed to higher redshift (say $z > 2$); these 
could be visible as SCUBA sources (Smail \etal~2002). Indeed, massive disks have recently been found at these redshifts (e.g., Labb{\' e} 
\etal~2003; Stockton, Canalizo, \& Maihara 2004). However, whether such disks had sufficient gas or were present in the requisite numbers to 
account for the formation of most massive ellipticals is still an open question. Our point here is not that gas-rich starbursts were uncommon 
in the early universe, but that these starbursts may have been driven by stochastic, ongoing merging of objects over a range of mass scales 
rather than ``simple" major mergers of fully-formed massive disks.  Other issues with the major merger scenario, such as specific frequencies, 
relative numbers of MP and MR GCs, and metallicities of the MR GCs, have been discussed in detail elsewhere (e.g., Ashman \& Zepf 1998; Harris 
2001).

It is clear that much larger samples of high S/N GC spectra will be needed to investigate the details of MR GC formation. Because
of downsizing, the most massive galaxies should have smaller age differences among their MR GCs but larger numbers of bright GCs to
probe these differences; less massive galaxies have fewer GCs but could have more age structure. Due to the steep central
concentration of GC systems, multi-object slit spectrographs, even if highly multiplexing, will continue to be limited in their
ability to observe large numbers of GCs. Fiber instruments on large telescopes or natural-seeing IFUs may be better suited to this
topic.

\section{Summary \& Conclusions}

We have undertaken a meta-analysis of high signal-to-noise Keck spectra of GCs in eight galaxies, ranging from dwarfs to massive ellipticals.
We have established the best constraints on GC ages that can realistically be achieved using current methods and technology. Ages have been
inferred from a direct comparison to Galactic GCs to avoid any dependence on stellar evolutionary synthesis models. We conclude that both the
MP and MR subpopulations in these galaxies are no younger than their Galactic counterparts ($\ga 10$ Gyr) and, within the uncertainties ($\sim
2$ Gyr), both subpopulations are coeval. These results indicate that the vast majority of GCs, at least in dense environments, were formed at
high redshift ($z > 2$). Given the expectation that GC formation has accompanied all major star forming episodes in a galaxy's history, the old
ages of GCs implies an old formation epoch for the stars in their host galaxies. This is consistent with results emerging from large
photometric surveys of galaxies, which are increasingly revealing massive galaxies already in place at similar redshifts.

We have synthesized a new GC formation scenario which contains elements of the in situ and accretion models that have been invoked in recent
years to explain the existence of GC subpopulations. The scenario suggests that MP GCs formed in low-mass halos in the early universe, with
their formation truncated by reionization. MR GCs formed along with the bulk of the field stars in massive early-type galaxies and spiral
bulges. This scenario qualitatively reproduces the MP and MR GC metallicity-galaxy mass relations and can explain the relative formation
efficiencies of MP and MR GCs in massive galaxies.

\acknowledgements

We thank Graeme Smith for useful comments and discussion. We acknowledge support by the National Science Foundation through Grant AST-0206139
and a Graduate Research Fellowship (J.S.). A.~J.~C.~acknowledges financial support from a UCM Fundaci\'on del Amo Fellowship. This research has
made use of the NASA/IPAC Extragalactic Database (NED) which is operated by the Jet Propulsion Laboratory, California Institute of Technology,
under contract with the National Aeronautics and Space Administration. The data presented herein were obtained at the W.M.~Keck Observatory,
which is operated as a scientific partnership among the California Institute of Technology, the University of California and the National
Aeronautics and Space Administration. The Observatory was made possible by the generous financial support of the W.M.~Keck Foundation. The
authors wish to recognize and acknowledge the very significant cultural role and reverence that the summit of Mauna Kea has always had within
the indigenous Hawaiian community. We are most fortunate to have the opportunity to conduct observations from this mountain.

\newpage

\epsfxsize=14cm
\epsfbox{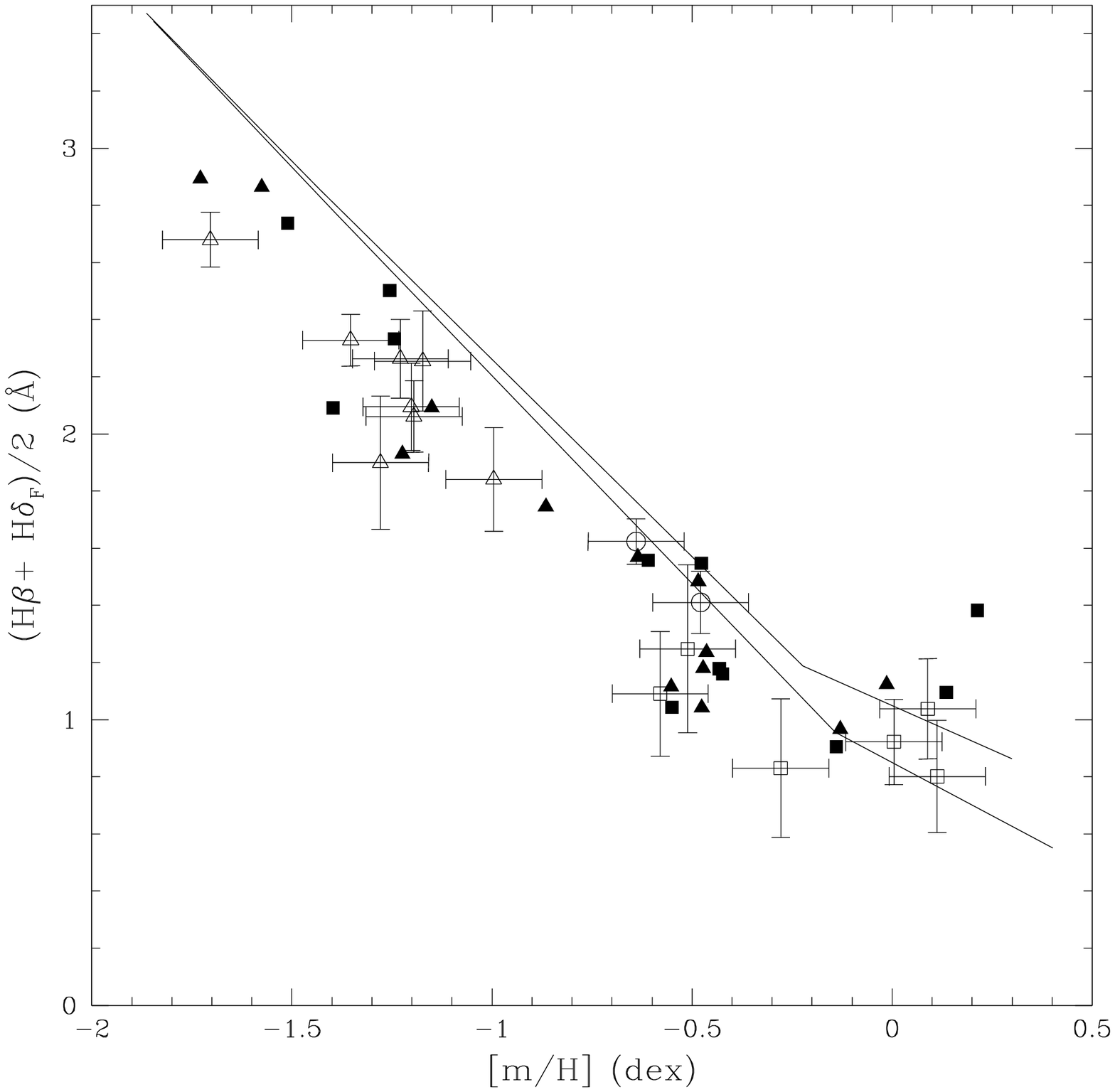} 
\figcaption[f1.eps]{\label{fig:fig1} Age-metallicity index-index plot for extragalactic GC subpopulations and Galactic GCs, with the mean of
H$\beta$ and H$\delta_F$ on the abscissa and PCA metallicity on the ordinate. The weighted subpopulations are plotted as open triangles
(metal-poor), open circles (intermediate-metallicity), and open squares (metal-rich). Individual Galactic GCs are plotted as filled triangles (Gregg
1994) and filled squares (Puzia \etal~2002b). At all metallicities, our subpopulations appear coeval with (or older than) the comparison Galactic
GCs. This suggests mean ages $\ga 10$--13 Gyr for our sample of extragalactic GCs. 14 Gyr model lines for [$Z$/H] = $-2.25$ to 0 and [$\alpha$/Fe] =
0 (bottom) and +0.3 (top) from Thomas, Maraston, \& Korn (2004) are superposed; these include a blue horizontal branch below [$Z$/H] = $-1.35$. At
fixed metallicity, older ages lie at weaker Balmer line strength. Since our data are not fully on the Lick system, we cannot directly compare them
to models. However, the models do show that our index combinations are relatively insensitive to [$\alpha$/Fe].}

\newpage

\begin{deluxetable}{lrrllll}
\rotate
\tablewidth{0pt}
\tabletypesize{\footnotesize}
\tablecaption{Galaxy Data
        \label{tab:pcares}}
\tablehead{Galaxy & Type & $M_V$\tablenotemark{a} & Subpopulations\tablenotemark{b} & No.~GCs\tablenotemark{c} & $m-M$ Reference\tablenotemark{d} & Spectra Reference\tablenotemark{e}}

\startdata

NGC 1407     & E0   & $-22.5$ & MP, IM, MR &	4, 9, 8	& Jensen, Tonry, \& Luppino (1998) & Cenarro \etal~(2005) \\
NGC 4365     & E3   & $-22.1$ & MP, IM, MR &	8, 9, 5 & Tonry \etal~(2001)   & Brodie \etal~(2005) \\
NGC 4594     & Sa   & $-22.1$ & MP, MR     &	7, 7 	& Tonry \etal~(2001)   & Larsen \etal~(2002) \\
NGC 3610     & E5   & $-21.5$ & MP, MR     &	3, 7 	& Tully (1988) & Strader, Brodie, \& Forbes (2004a)\\
NGC 1052     & E4   & $-21.0$ & MP, MR     &	7, 2	& Tonry \etal~(2001) & Pierce \etal~(2005) \\
NGC 7457     & S0   & $-19.6$ & MP, MR     &	9, 3	& Tonry \etal~(2001)   & Chomiuk, Strader, \& Brodie (2005) \\
VCC 1087     & dE,N & $-17.9$ & MP         &	9	& \nodata & Beasley \etal~(2005) \\
Fornax       & dSph & $-13.2$ & MP         &	4	& \nodata & Strader \etal~(2003b) \\

\enddata

\tablenotetext{a}{The absolute magnitudes are calculated using distance moduli from references listed in this table, $V$ photometry from
de Vaucouleurs \etal~(1991), and reddening corrections from Schlegel, Finkbeiner, \& Davis (1998), 
excepting the Fornax dSph and VCC 1087, whose $M_V$ come from Mateo (1998) and Jerjen, Binggeli, \& Barazza (2004), respectively.}
\tablenotetext{b}{Lists which of metal-poor (MP), intermediate-metallicity (IM), or metal-rich (MR) GC subpopulations are present.}
\tablenotetext{c}{Number of GCs in each subpopulation.}
\tablenotetext{d}{Literature reference for each distance modulus.}
\tablenotetext{e}{Literature reference for the original globular cluster spectra.}

\end{deluxetable}

\begin{deluxetable}{lcccccccccccccccccccccc}
\rotate
\setlength{\tabcolsep}{0.017in}
\tabletypesize{\scriptsize}
\tablecaption{Lick indices and PCA [m/H] of globular cluster subpopulations.\label{tbl-2}}
\tablewidth{0pt} 
\tablehead{
\colhead{Galaxy} &
\colhead{Subpop.} &   
\colhead{H$\delta_A$} &
\colhead{H$\delta_F$} &
\colhead{CN$_1$} &
\colhead{CN$_2$} &
\colhead{Ca4227} &
\colhead{G4300} &
\colhead{H$\gamma_A$} &
\colhead{H$\gamma_F$} &
\colhead{Fe4383} &
\colhead{Ca4455} &
\colhead{Fe4531} &
\colhead{Fe4668} &
\colhead{H$\beta$} &
\colhead{Fe5015} &
\colhead{Mg1} &
\colhead{Mg2} &
\colhead{Mg$b$} &
\colhead{Fe5270} &
\colhead{Fe5335} &
\colhead{Fe5406} &
\colhead{[m/H]}\\
\colhead{} &
\colhead{} &
\colhead{(\AA)} &
\colhead{(\AA)} &
\colhead{(mag)} &
\colhead{(mag)} &
\colhead{(\AA)} &
\colhead{(\AA)} &
\colhead{(\AA)} &
\colhead{(\AA)} &
\colhead{(\AA)} &
\colhead{(\AA)} &
\colhead{(\AA)} &
\colhead{(\AA)} &
\colhead{(\AA)} &
\colhead{(\AA)} &
\colhead{(mag)} &
\colhead{(mag)} &
\colhead{(\AA)} &
\colhead{(\AA)} &
\colhead{(\AA)} &
\colhead{(\AA)} &
\colhead{(dex)}}

\startdata

NGC 1407 & MP & 2.53  & 2.10  & $-0.044$ & $-0.016$ & 0.59 & 3.05 & $-0.54$ & 1.34  & 1.54 & 0.60 & 1.84 & 0.50 & 2.41 & 2.65 & 0.024  & 0.082 & 1.27 & 1.25 & 1.15 & 0.67 & $-1.17$ \\
NGC 4365 & MP & 1.31  & 1.46  & $-0.014$ & 0.017  & 0.47 & 3.49   & $-1.50$ & 0.80  & 2.02 & 0.62 & 1.90 & 1.34 & 2.22 & 2.86 & 0.042  & 0.111 & 1.75 & 1.30 & 1.33 & 0.89 & $-1.00$ \\
NGC 4594 & MP & 2.35  & 2.04  & $-0.037$ & $-0.010$ & 0.57 & 3.40 & $-0.01$ & 1.27  & 1.08 & 0.43 & 1.82 & 0.86 & 2.15 & 2.39 & $-0.008$ & 0.037 & 1.18 & 1.15 & 1.25 & 0.62 & $-1.20$ \\
NGC 3610 & MP & 0.27  & 1.51  & $-0.008$ & 0.018  & 0.14 & 3.09   & $-0.37$ & 1.55  & 1.66 & 0.17 & 1.81 & 1.30 & 2.29 & 2.02 & 0.017  & 0.051 & 0.70 & 0.82 & 1.33 & 0.46 & $-1.28$ \\
NGC 1052 & MP & 2.06  & 2.13  & $-0.021$ & $-0.001$ & 0.09 & 3.11 & $-0.66$ & 0.82  & 0.92 & 0.87 & 1.86 & 0.74 & 2.39 & 2.80 & 0.051  & 0.094 & 1.68 & 1.61 & 1.07 & 0.95 & $-1.23$ \\
NGC 7457 & MP & 2.20  & 1.95  & $-0.082$ & $-0.051$ & 0.59 & 3.00 & $-0.54$ & 1.03  & 0.82 & 0.62 & 1.25 & 0.59 & 2.17 & 2.14 & 0.036  & 0.083 & 0.93 & 1.47 & 1.16 & 0.69 & $-1.20$ \\
VCC 1087 & MP & 2.66  & 2.31  & $-0.065$ & $-0.039$ & 0.25 & 2.46 & $-0.08$ & 1.57  & 0.95 & 0.42 & 1.12 & 0.44 & 2.35 & 1.86 & 0.029  & 0.077 & 0.91 & 1.24 & 1.10 & 0.43 & $-1.35$ \\
Fornax   & MP & 4.00  & 2.80  & $-0.066$ & $-0.026$ & 0.19 & 1.23 & 2.14  & 2.35  & 0.41 & 0.15 & 0.87 & 0.01 & 2.56 & 0.85 & $-0.002$ & 0.020 & 0.54 & 0.62 & 0.61 & 0.34 & $-1.70$ \\
NGC 1407 & MR & $-1.66$ & 0.36  & 0.117  & 0.158  & 1.07 & 5.16   & $-5.71$ & $-1.13$ & 4.17 & 1.29 & 3.06 & 3.08 & 1.72 & 4.31 & 0.091  & 0.235 & 4.21 & 2.39 & 2.28 & 1.49 & $0.09$  \\
NGC 4365 & MR & $-1.65$ & 0.18  & 0.095  & 0.131  & 0.97 & 4.85   & $-4.85$ & $-0.91$ & 4.32 & 1.13 & 3.02 & 3.72 & 1.67 & 4.93 & 0.091  & 0.236 & 4.05 & 2.73 & 1.87 & 1.50 & $0.00$  \\
NGC 4594 & MR & $-2.45$ & $-0.08$ & 0.118  & 0.155  & 1.26 & 5.00 & $-5.68$ & $-1.02$ & 4.66 & 1.19 & 3.43 & 3.79 & 1.69 & 3.93 & 0.033  & 0.159 & 3.69 & 2.62 & 2.53 & 1.66 & $0.11$  \\
NGC 3610 & MR & 1.03  & 0.34  & 0.007  & 0.031  & 0.82 & 3.77     & $-2.75$ & 0.32  & 3.82 & 1.64 & 3.82 & 3.19 & 1.32 & 5.38 & 0.055  & 0.158 & 3.00 & 2.58 & 2.65 & 1.46 & $-0.28$ \\
NGC 1052 & MR & $-1.49$ & 0.06  & 0.007  & 0.024  & 0.98 & 3.14   & $-1.20$ & 0.49  & 2.48 & 0.79 & 2.23 & 2.12 & 2.43 & 3.50 & 0.069  & 0.203 & 2.80 & 2.71 & 1.60 & 0.86 & $-0.51$ \\
NGC 7457 & MR & $-0.86$ & 0.23  & 0.016  & 0.028  & 0.87 & 4.12   & $-3.15$ & $-0.10$ & 2.68 & 1.13 & 1.87 & 1.96 & 1.95 & 3.22 & 0.071  & 0.156 & 2.27 & 1.72 & 1.72 & 1.21 & $-0.58$ \\
NGC 1407 & IM & 0.63  & 1.27  & 0.020  & 0.056  & 0.61 & 4.24     & $-3.02$ & 0.18  & 2.97 & 1.04 & 2.48 & 1.60 & 1.97 & 3.97 & 0.047  & 0.143 & 2.52 & 1.80 & 1.59 & 1.05 & $-0.64$ \\
NGC 4365 & IM & $-0.24$ & 0.83  & 0.054  & 0.086  & 0.70 & 4.23   & $-3.30$ & 0.07  & 3.32 & 0.91 & 2.57 & 2.48 & 1.99 & 3.86 & 0.067  & 0.177 & 2.96 & 1.96 & 1.64 & 1.20 & $-0.48$ \\

\enddata
\end{deluxetable}

\begin{deluxetable}{lccccccccccccccccccccc}
\setlength{\tabcolsep}{0.018in}
\rotate
\tabletypesize{\scriptsize}
\tablecaption{Lick index weighted standard errors of the mean.\label{tbl-2a}}
\tablewidth{0pt}
\tablehead{
\colhead{Galaxy} &
\colhead{Subpop.} &
\colhead{H$\delta_A$} &
\colhead{H$\delta_F$} &
\colhead{CN$_1$} &
\colhead{CN$_2$} &
\colhead{Ca4227} &
\colhead{G4300} & 
\colhead{H$\gamma_A$} &
\colhead{H$\gamma_F$} &
\colhead{Fe4383} &
\colhead{Ca4455} &  
\colhead{Fe4531} &
\colhead{Fe4668} &
\colhead{H$\beta$} &
\colhead{Fe5015} &
\colhead{Mg1} &   
\colhead{Mg2} &   
\colhead{Mg$b$} & 
\colhead{Fe5270} &
\colhead{Fe5335} &
\colhead{Fe5406}\\
\colhead{} &
\colhead{} &
\colhead{(\AA)} &
\colhead{(\AA)} &
\colhead{(mag)} &
\colhead{(mag)} &
\colhead{(\AA)} &
\colhead{(\AA)} &
\colhead{(\AA)} &
\colhead{(\AA)} &
\colhead{(\AA)} &
\colhead{(\AA)} &
\colhead{(\AA)} &
\colhead{(\AA)} &
\colhead{(\AA)} &
\colhead{(\AA)} &
\colhead{(mag)} &
\colhead{(mag)} &
\colhead{(\AA)} & 
\colhead{(\AA)} &
\colhead{(\AA)} &
\colhead{(\AA)}}
\startdata

NGC 1407 & MP & 0.18 & 0.13 & 0.005 & 0.007 & 0.10 & 0.17 & 0.18 & 0.11 & 0.26 & 0.13 & 0.20 & 0.30 & 0.12 & 0.26 & 0.003 & 0.003 & 0.13 & 0.15 & 0.17 & 0.13 \\
NGC 4365 & MP & 0.15 & 0.12 & 0.006 & 0.007 & 0.08 & 0.20 & 0.44 & 0.17 & 0.21 & 0.10 & 0.16 & 0.23 & 0.14 & 0.21 & 0.003 & 0.003 & 0.08 & 0.05 & 0.12 & 0.11 \\
NGC 4594 & MP & 0.14 & 0.09 & 0.004 & 0.005 & 0.08 & 0.13 & 0.14 & 0.09 & 0.21 & 0.10 & 0.20 & 0.31 & 0.12 & 0.25 & 0.003 & 0.003 & 0.14 & 0.13 & 0.16 & 0.12 \\
NGC 3610 & MP & 0.29 & 0.18 & 0.008 & 0.010 & 0.16 & 0.25 & 0.26 & 0.15 & 0.41 & 0.19 & 0.28 & 0.40 & 0.15 & 0.34 & 0.003 & 0.004 & 0.16 & 0.17 & 0.21 & 0.15 \\
NGC 1052 & MP & 0.14 & 0.09 & 0.004 & 0.005 & 0.08 & 0.14 & 0.14 & 0.09 & 0.22 & 0.11 & 0.17 & 0.27 & 0.10 & 0.23 & 0.002 & 0.003 & 0.11 & 0.12 & 0.13 & 0.10 \\
NGC 7457 & MP & 0.13 & 0.09 & 0.004 & 0.005 & 0.07 & 0.13 & 0.13 & 0.08 & 0.19 & 0.10 & 0.15 & 0.23 & 0.09 & 0.20 & 0.002 & 0.002 & 0.10 & 0.11 & 0.13 & 0.09 \\
VCC 1087 & MP & 0.09 & 0.06 & 0.003 & 0.003 & 0.05 & 0.09 & 0.10 & 0.06 & 0.16 & 0.07 & 0.11 & 0.18 & 0.06 & 0.14 & 0.002 & 0.002 & 0.07 & 0.08 & 0.09 & 0.07 \\
Fornax   & MP & 0.11 & 0.07 & 0.003 & 0.004 & 0.06 & 0.10 & 0.10 & 0.06 & 0.15 & 0.07 & 0.11 & 0.17 & 0.06 & 0.13 & 0.001 & 0.002 & 0.06 & 0.07 & 0.08 & 0.06 \\
NGC 1407 & MR & 0.22 & 0.14 & 0.006 & 0.007 & 0.10 & 0.16 & 0.19 & 0.11 & 0.22 & 0.12 & 0.17 & 0.26 & 0.10 & 0.21 & 0.002 & 0.003 & 0.11 & 0.12 & 0.13 & 0.10 \\
NGC 4365 & MR & 0.17 & 0.12 & 0.006 & 0.006 & 0.07 & 0.14 & 0.17 & 0.10 & 0.18 & 0.08 & 0.13 & 0.25 & 0.09 & 0.19 & 0.003 & 0.004 & 0.10 & 0.11 & 0.15 & 0.09 \\
NGC 4594 & MR & 0.18 & 0.13 & 0.005 & 0.006 & 0.09 & 0.16 & 0.18 & 0.10 & 0.23 & 0.13 & 0.21 & 0.34 & 0.15 & 0.29 & 0.003 & 0.004 & 0.15 & 0.16 & 0.20 & 0.13 \\
NGC 3610 & MR & 0.26 & 0.19 & 0.008 & 0.009 & 0.15 & 0.27 & 0.27 & 0.16 & 0.35 & 0.17 & 0.22 & 0.41 & 0.15 & 0.30 & 0.003 & 0.005 & 0.17 & 0.17 & 0.20 & 0.15 \\
NGC 1052 & MR & 0.31 & 0.21 & 0.009 & 0.010 & 0.16 & 0.29 & 0.32 & 0.20 & 0.45 & 0.23 & 0.36 & 0.54 & 0.21 & 0.45 & 0.005 & 0.005 & 0.21 & 0.23 & 0.26 & 0.19 \\
NGC 7457 & MR & 0.25 & 0.17 & 0.007 & 0.008 & 0.12 & 0.21 & 0.24 & 0.14 & 0.31 & 0.15 & 0.25 & 0.36 & 0.13 & 0.29 & 0.003 & 0.004 & 0.15 & 0.17 & 0.19 & 0.14 \\
NGC 1407 & IM & 0.09 & 0.06 & 0.003 & 0.003 & 0.04 & 0.08 & 0.08 & 0.05 & 0.11 & 0.06 & 0.08 & 0.13 & 0.05 & 0.11 & 0.001 & 0.001 & 0.06 & 0.06 & 0.07 & 0.05 \\
NGC 4365 & IM & 0.11 & 0.08 & 0.003 & 0.003 & 0.05 & 0.10 & 0.13 & 0.08 & 0.14 & 0.07 & 0.12 & 0.18 & 0.08 & 0.17 & 0.003 & 0.003 & 0.08 & 0.10 & 0.10 & 0.07 \\

\enddata
\end{deluxetable}

\begin{deluxetable}{lccccccccccccccccccccc}
\setlength{\tabcolsep}{0.018in}
\rotate
\tabletypesize{\scriptsize}
\tablecaption{Lick index weighted standard deviations.\label{tbl-3a}}
\tablewidth{0pt} 
\tablehead{
\colhead{Galaxy} &
\colhead{Subpop.} &
\colhead{H$\delta_A$} &
\colhead{H$\delta_F$} &
\colhead{CN$_1$} &
\colhead{CN$_2$} &
\colhead{Ca4227} &
\colhead{G4300} &
\colhead{H$\gamma_A$} &
\colhead{H$\gamma_F$} &
\colhead{Fe4383} &
\colhead{Ca4455} &
\colhead{Fe4531} &
\colhead{Fe4668} &
\colhead{H$\beta$} &
\colhead{Fe5015} &
\colhead{Mg1} &
\colhead{Mg2} &
\colhead{Mg$b$} &
\colhead{Fe5270} &
\colhead{Fe5335} &
\colhead{Fe5406}\\
\colhead{} &
\colhead{} &
\colhead{(\AA)} &
\colhead{(\AA)} &
\colhead{(mag)} &
\colhead{(mag)} &
\colhead{(\AA)} &
\colhead{(\AA)} &
\colhead{(\AA)} &
\colhead{(\AA)} &
\colhead{(\AA)} &
\colhead{(\AA)} &
\colhead{(\AA)} &
\colhead{(\AA)} &
\colhead{(\AA)} &
\colhead{(\AA)} &
\colhead{(mag)} &
\colhead{(mag)} &
\colhead{(\AA)} &
\colhead{(\AA)} &
\colhead{(\AA)} &
\colhead{(\AA)}}
\startdata

NGC 1407 & MP & 0.67 & 0.33 & 0.028 & 0.017 & 0.39 & 0.65 & 1.15 & 0.53 & 1.08 & 0.25 & 0.32 & 1.32 & 0.11 & 1.08 & 0.015 & 0.026 & 0.72 & 0.14 & 0.18 & 0.42 \\
NGC 4365 & MP & 0.34 & 0.27 & 0.017 & 0.021 & 0.24 & 0.41 & 1.12 & 0.43 & 0.64 & 0.27 & 0.40 & 0.54 & 0.29 & 0.39 & 0.009 & 0.007 & 0.17 & 0.10 & 0.41 & 0.31 \\
NGC 4594 & MP & 0.27 & 0.36 & 0.012 & 0.017 & 0.41 & 1.08 & 1.51 & 0.58 & 0.69 & 0.28 & 0.95 & 0.64 & 0.60 & 1.99 & 0.020 & 0.033 & 0.39 & 0.41 & 0.63 & 0.59 \\
NGC 3610 & MP & 1.48 & 0.21 & 0.045 & 0.045 & 0.17 & 0.66 & 0.81 & 0.46 & 1.02 & 0.13 & 1.05 & 1.60 & 0.27 & 2.04 & 0.015 & 0.012 & 0.48 & 0.54 & 0.35 & 1.08 \\
NGC 1052 & MP & 0.95 & 0.67 & 0.069 & 0.061 & 0.45 & 1.16 & 1.51 & 0.47 & 2.16 & 0.37 & 0.67 & 1.78 & 0.57 & 1.77 & 0.038 & 0.026 & 0.99 & 0.91 & 1.10 & 0.32 \\
NGC 7457 & MP & 0.82 & 0.48 & 0.014 & 0.024 & 0.36 & 0.75 & 0.94 & 0.35 & 1.02 & 0.38 & 0.56 & 0.68 & 0.31 & 1.26 & 0.016 & 0.025 & 0.58 & 0.43 & 0.46 & 0.59 \\
VCC 1087 & MP & 1.08 & 0.38 & 0.031 & 0.027 & 0.26 & 1.03 & 0.54 & 0.48 & 0.82 & 0.25 & 0.82 & 0.70 & 0.29 & 0.54 & 0.012 & 0.029 & 0.49 & 0.27 & 0.18 & 0.19 \\
Fornax   & MP & 0.91 & 0.59 & 0.016 & 0.012 & 0.07 & 0.79 & 1.40 & 0.71 & 0.85 & 0.06 & 0.51 & 0.41 & 0.20 & 0.60 & 0.002 & 0.012 & 0.23 & 0.30 & 0.18 & 0.14 \\
NGC 1407 & MR & 0.65 & 0.35 & 0.037 & 0.040 & 0.52 & 0.59 & 0.91 & 0.26 & 0.60 & 0.23 & 0.38 & 0.72 & 0.41 & 1.13 & 0.024 & 0.024 & 0.49 & 0.28 & 0.35 & 0.21 \\
NGC 4365 & MR & 0.54 & 0.40 & 0.032 & 0.036 & 0.16 & 0.44 & 0.71 & 0.41 & 0.56 & 0.19 & 0.44 & 0.66 & 0.26 & 0.65 & 0.009 & 0.012 & 0.20 & 0.25 & 0.21 & 0.18 \\
NGC 4594 & MR & 2.01 & 0.61 & 0.086 & 0.083 & 0.68 & 0.89 & 1.26 & 0.60 & 1.34 & 0.43 & 0.57 & 2.08 & 0.38 & 1.07 & 0.019 & 0.040 & 0.69 & 0.88 & 0.63 & 0.95 \\
NGC 3610 & MR & 2.06 & 2.42 & 0.029 & 0.067 & 1.15 & 1.16 & 1.28 & 0.87 & 1.56 & 0.95 & 2.13 & 2.11 & 0.29 & 0.96 & 0.023 & 0.019 & 1.03 & 0.24 & 0.89 & 0.47 \\
NGC 7457 & MR & 0.60 & 0.71 & 0.010 & 0.020 & 0.12 & 0.50 & 0.22 & 0.24 & 0.64 & 0.05 & 0.26 & 0.54 & 0.14 & 0.31 & 0.005 & 0.011 & 0.25 & 0.66 & 0.67 & 0.25 \\
NGC 1407 & IM & 0.87 & 0.38 & 0.038 & 0.039 & 0.15 & 0.50 & 0.69 & 0.41 & 0.40 & 0.26 & 0.64 & 0.64 & 0.20 & 0.52 & 0.013 & 0.027 & 0.48 & 0.26 & 0.21 & 0.27 \\
NGC 4365 & IM & 0.22 & 0.17 & 0.013 & 0.016 & 0.23 & 0.30 & 0.37 & 0.21 & 0.41 & 0.15 & 0.41 & 0.34 & 0.26 & 0.73 & 0.012 & 0.013 & 0.29 & 0.23 & 0.09 & 0.16 \\

\enddata
\end{deluxetable}

\end{document}